 \documentstyle[12pt,twoside,fleqn,espcrc1,epsf]{article}
\newcommand{\AmS}{{\protect\the\textfont2
  A\kern-.1667em\lower.5ex\hbox{M}\kern-.125emS}}

\title{Structures and Transitions in Light Unstable Nuclei
}

\author{Y. Kanada-En'yo\address{Institute of Particle and Nuclear Studies,\\ 
High Energy Accelerator Research Organization, \\
Oho 1-1, Tsukuba-shi 305-0801, Japan
}
, H. Horiuchi\address{Department of Physics,\\ Kyoto University,\\ 
Kitashirakawa-Oiwake, Sakyo-ku, \\
Kyoto 606-01, 
Japan
}
 and A, Dot\'e$^{\rm b}$
}

\begin{document}

% typeset front matter
\maketitle

\begin{abstract}
  We study the structures of the unstable Be isotopes with the 
theoretical method of antisymmetrized molecular dynamics.
It is found that various structures of the excited states appear in 
the low-energy region of neutron-rich Be nuclei. 
Focusing on the 2$\alpha$ clustering, 
we analyze the intrinsic structures with the help of the 
experimental data of Gamow-Teller transitions.
\end{abstract}

\section{INTRODUCTION}
Recently, the study of unstable nuclei is one of the hot subjects.
In the light unstable nuclei, the experimental informations 
near the drip lines increase rapidly and 
many interesting phenomena are found.
The exotic structures such as the 
halo, the skin structures, and the clustering structures 
are suggested in some light unstable nuclei.
Are the features of the clusters in unstable nuclei
similar to those in the stable nuclei ?
The shift of the magic number and 
the deformations of unstable nuclei are the concerned problems.

Our aim is to research on the excited states of unstable 
and to study systematically the various structures of unstable nuclei.
We apply a theoretical method of antisymmetrized molecular dynamics
(AMD). The method of  variation after projection(VAP) in the framework
of AMD has been already proved to be useful to investigate the 
excited states of light unstable nuclei \cite{ENYOf,ENYOg}
as well as stable nuclei \cite{ENYOe}. 
AMD is a powerful approach which is applicable to 
the excited states of general nuclei.
The wave function of AMD can represent various clustering structures
as well as shell-model-like structures, because there is no inert core
in the framework.

In the previous study on the excited states of 
$^{10}$Be with AMD \cite{ENYOg,DOTE},
we found that the molecule-like structures may coexist
in the low-energy region with the ordinary shell-model-like states.
The data of the transitions such as $E2$ and Gamow-Teller transitions
are so helpful to classify the excited levels into rotational bands.
The reader is referred to 
the paper \cite{ENYOg} for the detailed discussions.

\section{FORMULATION}

In this section we explain the formulation of AMD for the study
of the nuclear structure of the excited states.

\subsection{Wave function}
The wave function of a system is written by AMD wave functions,
\begin{equation}
\Phi=c \Phi_{AMD} +c' \Phi '_{AMD} + \cdots .
\end{equation}
An AMD wavefunction of a nucleus with mass number $A$
is a Slater determinant of Gaussian wave packets;
\begin{eqnarray}
&\Phi_{AMD}({\bf Z})=\frac{1}{\sqrt{A!}}
{\cal A}\{\varphi_1,\varphi_2,\cdots,\varphi_A\},\\
&\varphi_i=\phi_{{\bf X}_i}\chi_{\xi_i}\tau_i :\left\lbrace
\begin{array}{l}
\phi_{{\bf X}_i}({\bf r}_j) \propto
\exp\left 
[-\nu\biggl({\bf r}_j-\frac{{\bf X}_i}{\sqrt{\nu}}\biggr)^2\right],\\
\chi_{\xi_i}=
\left(\begin{array}{l}
{1\over 2}+\xi_{i}\\
{1\over 2}-\xi_{i}
\end{array}\right),
\end{array}\right. 
\end{eqnarray}
where the $i$th single-particle wave function $\varphi_i$
is a product of the spatial wave function, the intrinsic spin function and 
the iso-spin function. The spatial part $\phi_{{\bf X}_i}$ is presented by 
complex parameters $X_{1i}$, $X_{2i}$, $X_{3i}$, of the centers of 
Gaussians, and $\chi_{\xi_i}$ is the intrinsic spin function parameterized by
$\xi_{i}$. $\tau_i$ is the iso-spin
function which is fixed to be up(proton) or down(neutron)
in the present calculations.
Thus an AMD wave function is parameterized by a set of complex parameters
${\bf Z}\equiv \{X_{ni},\xi_i\}\ (n=1,3\ \hbox{and }  i=1,A)$
which determine the centers of 
Gaussians of the spatial parts
and the directions of the intrinsic spins for all single-particle wave 
functions.

If we consider a spin-parity eigen state projected from a AMD wave function,
the total wave function is linear combinations 
of Slater determinants,
\begin{equation}
\Phi({\bf Z})=P^{J\pm}_{MK'}\Phi_{AMD}({\bf Z}) = (1\pm P)
\int d\Omega D^{J*}_{MK'}(\Omega)R(\Omega)\Phi_{AMD}({\bf Z}),
\end{equation}
where $P$ is a parity projection operator.
The integrations for the Euler angles are numerically  calculated by 
a summation of mesh points on $\Omega$. 

In principal the total wave function can be a superposition of independent
AMD wave functions, 
\begin{equation}
\Phi=cP^{J\pm}_{MK'}\Phi_{AMD}({\bf Z})
+c'P^{J\pm}_{MK'}\Phi_{AMD}({\bf Z}')+\cdots.
\end{equation}

\subsection{Energy variation}
We make variational calculations
to find the state which minimizes the energy of the system;
$\langle\Phi|H|\Phi\rangle/\langle\Phi|\Phi\rangle$
by the method of frictional cooling \cite{ENYObc,ENYOa}, one of the 
imaginary time methods.
The time development of the wave function $\Phi({\bf Z})$ is given
by the equations,
\begin{equation}
\frac{dX_{n k}}{dt}=
(\lambda+i\mu)\frac{1}{i \hbar} \frac{\partial}{\partial X^*_{n k} }
\frac{\langle \Phi({\bf Z})|H|\Phi({\bf Z})\rangle}{\langle \Phi({\bf Z})
|\Phi({\bf Z})\rangle},
\quad (n=1,3\quad k=1,A)
\end{equation}
\begin{equation}
\frac{d\xi_{k}}{dt}=(\lambda+i\mu)\frac{1}{i\hbar}
\frac{\partial}{\partial\xi^*_{k}}
\frac{\langle \Phi({\bf Z})|H|\Phi({\bf Z})\rangle}{\langle \Phi({\bf Z})
|\Phi({\bf Z})\rangle},
\quad (k=1,A)
\end{equation}
with arbitrary real numbers $\lambda$ and $\mu < 0$. It is easily proved that 
 the energy of the system decreases with time. After sufficient time
steps of cooling, we obtain
the optimum parameters for the
minimum-energy state.

\subsection{Lowest $J^\pm$ states}
In order to obtain the wave function for the lowest $J^\pm$ state,
we perform the energy variation for the spin parity eigenstates projected
from an AMD wave function. In this case the trial function is 
$\Phi=P^{J\pm}_{MK'}\Phi_{AMD}({\bf Z})$. 
For the preparation of VAP calculations, first
we make variational calculations before 
the spin projection
in order to choose an initial wave function for each parity and 
an appropriate $K'$ quantum for each spin-parity state.
Then we perform VAP calculation 
for 
\begin{equation}\nonumber
\langle P^{J\pm}_{MK'}\Phi_{AMD}({\bf Z})|H|P^{J\pm}_{MK'}
\Phi_{AMD}({\bf Z}) \rangle/ 
\langle P^{J\pm}_{MK'}({\bf Z})|P^{J\pm}_{MK'}({\bf Z})
 \rangle
\end{equation}
 with the adopted $K'$ quantum from the initial state.
In the VAP procedure, the $3$-axis of Euler angle in the total-angular 
momentum projection is not necessarily same as the principal $z$-axis.
The direction of the intrinsic deformation in the body-fixed frame 
is automatically determined in the energy variation depending on the 
adopted $K'$ quantum.
However, in many cases, the approximately principal axis $z$
 obtained by a VAP calculation is found to be almost equal to the $3$-axis.

\subsection{Higher excited states \label{subsec:excited}}
As mentioned above,
with the VAP calculation for 
$\Phi({\bf Z})=P^{J\pm}_{MK'}\Phi_{AMD}({\bf Z})$,
we obtain the set of parameters ${\bf Z}={\bf Z}^{J\pm}_1$ which presents
the wave function for the first $J^\pm$ state. 
To search the parameters ${\bf Z}$  for the higher excited $J^\pm$ states,
we make the energy variation for the orthogonal component 
of a spin-parity projected AMD wave function
to the lower states by superposing the wave functions.
If the AMD wave functions for the higher excited states
have energetically local minimums, we do not need orthogonalization 
in VAP calculations. 

\subsection{Expectation values \label{subsec:expect}}  

After VAP calculations for various $J^\pm_n$ states,  we obtain many 
intrinsic states
$\Phi^1_{AMD}$, 
$\Phi^2_{AMD}$,$\cdots$, 
$\Phi^m_{AMD}$.
Finally we improve the wave functions
by diagonalizing the Hamiltonian matrix 
$\langle P^{J\pm}_{MK'} \Phi^i_{AMD}
|H|P^{J\pm}_{MK''} \Phi^j_{AMD}\rangle$
and the norm matrix
$\langle P^{J\pm}_{MK'} \Phi^i_{AMD}
|P^{J\pm}_{MK''} \Phi^j_{AMD}\rangle$
simultaneously with regard to ($i,j$) for 
all the intrinsic states and ($K', K''$).
In comparison with the experimental data, 
the theoretical values are calculated with the 
final states after the diagonalization.

\section{RESULTS}

We calculate the excited states of $^{12}$Be and $^{11}$Be.
The adopted interactions are the MV1 case3 (the finite two-body and
the zero-range three-body interactions \cite{TOHSAKI}) for the central force,
and G3RS for the spin-orbit force \cite{LS}. 
The coulomb interaction is approximated 
by a sum of seven Gaussians. 
In the present calculations, we adopt the Majorana parameter $m=0.65$ 
and omit the Bartlett and Heisenberg terms 
in the two-body central force. As for the strength of the spin-orbit force, 
we try two cases of the parameter,  $u_{ls}$=3700 MeV and $u_{ls}$=2500 MeV.

The resonance states are treated as a bound-state approximation 
in AMD, because the restriction of the Gaussian forms of the wave functions 
makes an artificial barrier for outgoing particles.

\subsection{$^{12}$Be}
\begin{figure}
\caption{\label{fig:be12spe} Energy levels and transitions of $^{12}$Be. 
The adopted parameters for the calculations are Majorana parameter $m=0.65$ 
and the strength of the spin-orbit force $u_{ls}=3700$ MeV.  }

\centerline{\epsfxsize= 12cm\epsffile{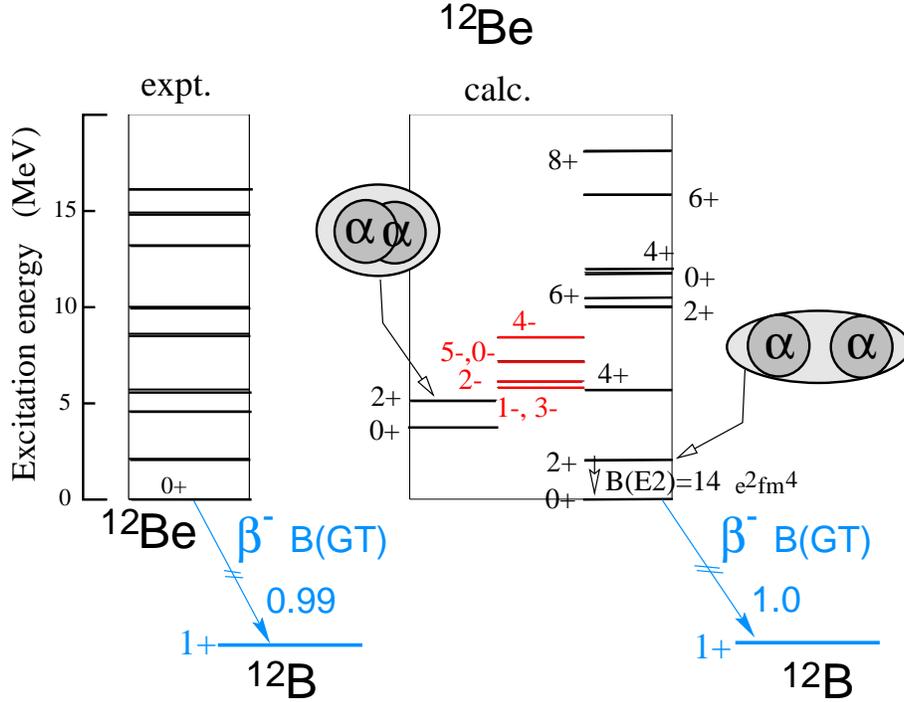}} 
\end{figure}

The energy levels of $^{12}$Be are displayed in 
Figure \ref{fig:be12spe}.
The theoretical results are calculated with 
the stronger spin-orbit force as $u_{ls}=3700$ MeV
which can reproduce the abnormal parity of the ground $1/2^+$ state
in $^{11}$Be. Many excited states are seen in the low-energy
region. Even though the $^{12}$Be has the neutron magic number 8, 
it is surprising that the calculated ground state is not the ordinary
state with the neutron closed $p$-shell, but a 2 particle 2 hole
state with the developed clustering structure.
As the evidence of the breaking of the closed $p$-shell, the calculations
reproduce well the small value of the Gamow-Teller(GT) transition 
strength from $^{12}$Be($0^+$) to $^{12}$B($1^+$). 
The experimental $B(GT)$ is about 
one third as small as the value expected for the closed 
$p$-shell $^{12}$Be($0^+$). 
In other words, the weak GT transition is one of the proofs
for the disappearance of the neutron magic number 8 in neutron-rich 
Be isotopes.
It is consistent with the shell model analysis by T. Suzuki et al. 
\cite{SUZUKIa}.
We also find many excited states with the developed molecule-like 
structures which construct the rotational bands $K^\pi=0^+_1$,
$K^\pi=1^-_1$, $K^\pi=0^+_3$ in the low-energy region.
Since the  ground state has 
the large deformation in the present calculations,
The predicted $B(E2; 2^+_1\rightarrow 0^+_1)$ in the ground band
is 14 e$^2$fm$^4$ and is much larger than the value 7 e$^2$fm$^4$ 
of $B(E2; 2^+_2\rightarrow 0^+_2)$ in the second 
$K^\pi=0^+$ band which consists of the closed $p$-shell states. 

\subsection{$^{11}$Be}

\begin{figure}
\caption{\label{fig:be11spe}
Energy levels of $^{11}$Be. 
The adopted parameter for the 
strength of the spin-orbit force is $u_{ls}=3700$ MeV for case(1)
and $u_{ls}=2500$ MeV for case(2).  }

\centerline{\epsfxsize 12cm\epsffile{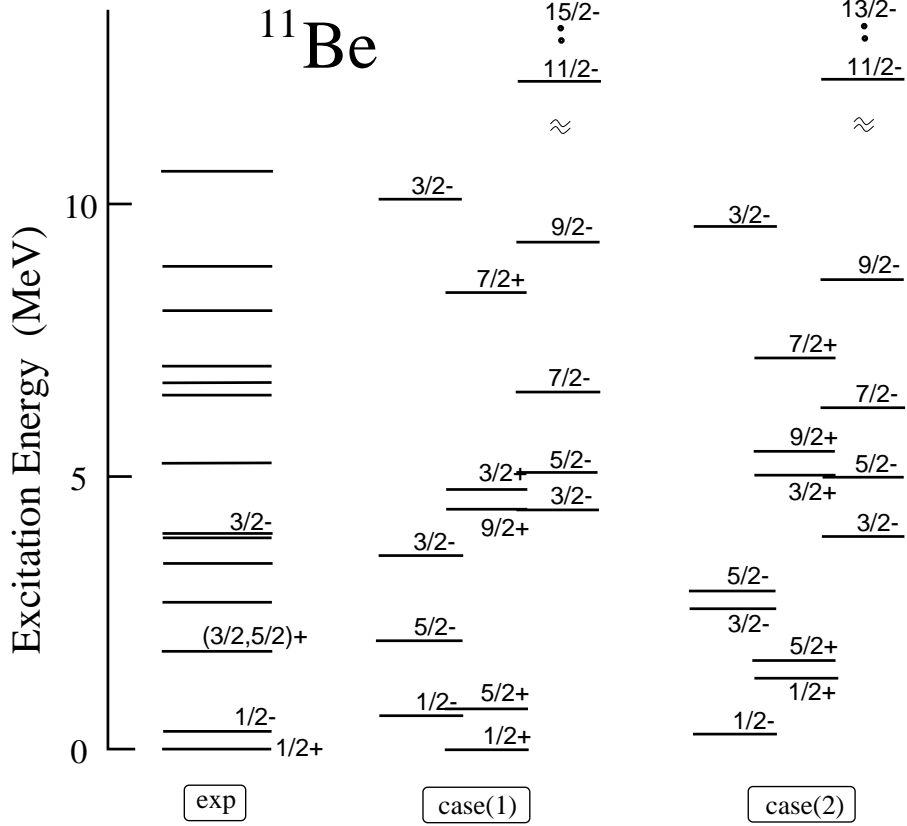} }
\end{figure}

In this subsection, we discuss the excited states of $^{11}$Be.
In Fig \ref{fig:be11spe}, the calculated energy levels are shown with 
the experimental data. We use the rather strong spin-orbit force in 
case (1) and the weaker LS force in case (2). The binding energies are 
underestimated as 58 MeV in case (1) and as 54 MeV in case (2).
It is not difficult to fit to the experimental binding energy 
by choosing a smaller value for the Majorana parameter.
 
In the case (1) with the 
stronger LS force, the abnormal spin-parity $1/2^+$ state comes down
lower than the normal negative parity states, which is consistent with 
the experimental data. 
In both cases, we found some rotational bands $K^\pi=1/2^-$,
$K^\pi=1/2^+$ and $K^\pi=3/2^-$, although the order of the levels 
depends on the adopted interaction.
The $K^\pi=3/2^-$ band has a large moment of inertia and reaches to 
the high spin states. The highest spin is $15/2^-$ in case(1) and 
$13/2^-$ in case (2). It is expected that 
the $K^\pi=3/2^-$ band may reach to the further high spin such as 
a $17/2^-$ state in the calculations with a smaller value of 
the Majorana parameter. 

\begin{figure}
\caption{\label{fig:be11int} 
Schematic figures for the intrinsic structures of $^{11}$Be. 
(a), (b), (c), (d) correspond to $K=1/2^-$, $K=1/2^+$, $K=3/2^-$,
$3/2^-(\approx 10$ MeV), respectively.}
\centerline{\epsfxsize 14cm\epsffile{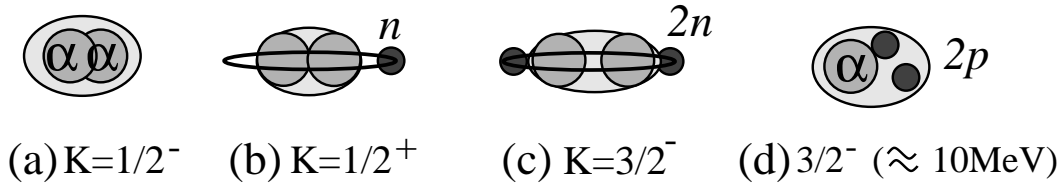}} 
\end{figure}

Next we analyze the intrinsic wave functions 
of the obtained rotational bands.
In the results of $^{11}$Be, 2$\alpha$ clustering structures appear very
often in many states. The schematic figures are presented in 
Fig. \ref{fig:be11int}.
Two alpha cores are found in the states of 
the lowest negative parity band $K^\pi=1/2^-$ 
(Fig.\ref{fig:be11int}(a)). 
In the positive parity
states, one neutron is in the $sd$ orbit and the 2$\alpha$ clustering
develops (Fig.\ref{fig:be11int}(b)). 
We think that one of the reason of the energy gain of the 
positive parity states is considered to be the clustering development.
The interesting point is that the $K^\pi=3/2^-$
band comes from the well-developed molecule-like structure
with 2 neutrons in the $sd$ orbit (Fig.\ref{fig:be11int}(c)).
The large moment of inertia results 
from the largely deformed intrinsic structure.
At about 10 MeV, we found another characteristic state where
one of the 2 $\alpha$-clusters completely breaks (Fig.\ref{fig:be11int}(d)). 
In this state the intrinsic spins of 2 protons couple up to be totally a unit.
As mentioned above, various structures coexist in the low energy region in 
$^{11}$Be. It is expected that the valence neutrons play important roles
in these excited states.

\begin{figure}
\caption{\label{fig:be11beta}
Beta decays to the excited states of  $^{11}$Be. 
GT transitions are calculated. }
\centerline{\epsfxsize 13cm\epsffile{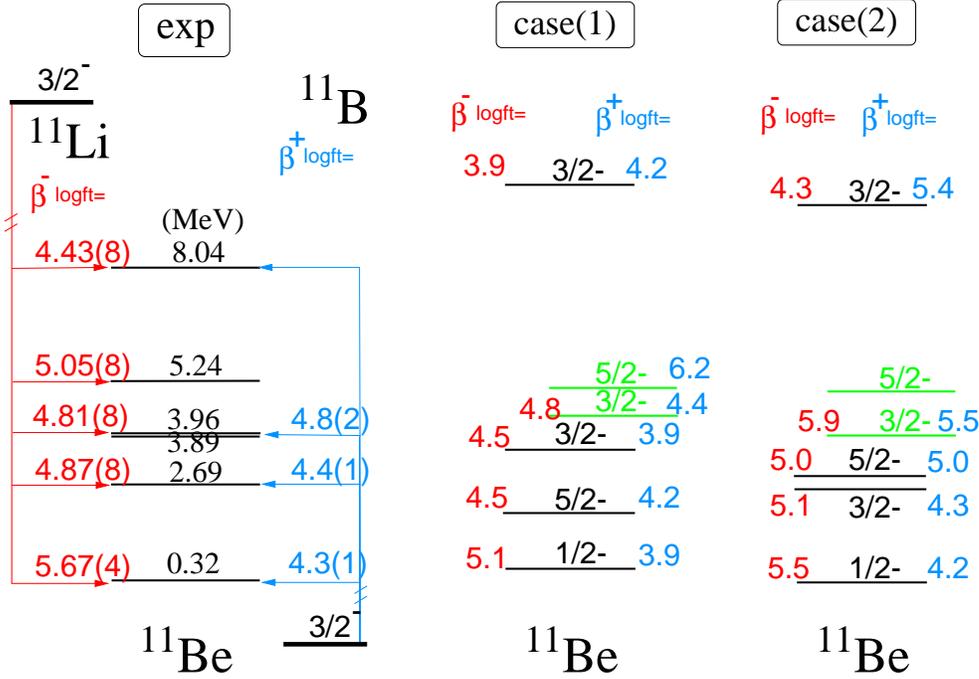}} 
\end{figure}

We discuss the Gamow-Teller(GT) transitions in order to analyze the 
level structures of $^{11}$Be. 
First we consider $\beta$ decay from $^{11}$B.
The experimental data of GT-type transitions from the ground state of 
$^{11}$B are deduced from the $0^\circ$ cross sections of the charge
exchange reactions \cite{FUJIWARA}.
Comparing the calculated log ft values with experimental data 
(Fig.\ref{fig:be11beta}),
the observed three levels well correspond to the excited states in the
lowest $K^\pi=1/2^-$ band. It is natural that the GT transitions from 
$^{11}$B to these levels are reasonably strong 
because both the parent and the daughter states are ordinary $p$-shell states.

The beta decays from $^{11}$Li are shown in Fig.\ref{fig:be11beta}.
We notice that the ideal 2$\alpha$-cluster structures 
in the daughter
$^{11}$Be states forbid any GT transitions from $^{11}$Li
because of the Pauli forbidden. In other words, the GT transitions from
$^{11}$Li are allowed only if the daughter states of $^{11}$Be
have the component with the breaking of the $\alpha$ clusters.
The calculations well agree with the experimental log ft values.
According to the calculated results, the lowest $3/2^-$ and $5/2^-$ states
in the $K^\pi=1/2^-$ band have the more than 10 \% breaking of the clusters.
One of the reason for the weak GT transition to the lowest $1/2^-$ state
is the clustering structure of $^{11}$Be.
Another reason is suggested by Suzuki et.al \cite{SUZUKIa}
as the halo structure
of the parent $^{11}$Li state, which may not be described enough in the
present AMD calculations. 
In the calculations, transitions 
for the higher excited states in the $K^\pi=3/2^-$ band
are suppressed because of the $\alpha$ clusters in the developed
molecule-like structures.
The strong transition to the excited state at 8 MeV well fits to
the calculated $3/2^-$ state ($\approx$10 MeV ) where one of the 
two $\alpha$ clusters completely breaks. 

As for the $E1$ transitions, the 
$B(E1;1/2^-\rightarrow 1/2^+)$ of the present calculations 
is much smaller than the experimental data.
 We should describe the wave function more
precisely to reproduce the strength $B(E1)$.

\section{SUMMARY}

We studied the structure of the excited states of $^{12}$Be and $^{11}$Be
and analyzed the level structure with the help of the data of
 Gamow-Teller transitions.

In $^{12}$Be, it is found that the ground state has the developed clustering
structure instead of the neutron $p$-shell closed state.
The main component of the ground state is the $2p$-$2h$ state in
terms of a simple shell model.
That is to say the neutron magic number 8 disappears in $^{12}$Be. 
It is consistent with the small GT strength to $^{12}$Be.
Many molecule-like states make rotational bands
in low energy regions.

In the case of $^{11}$Be, various structures are seen in the
excited states.
In the results rotational bands with the 
the developed clustering states are suggested.
We discuss the Gamow-Teller transitions comparing the AMD calculations 
with the experimental data.
The experimental data of GT transitions from $^{11}$B well agree with the 
theoretical results of the decay to the ordinary $p$-shell states
of $^{11}$Be.
With the help of the log ft values for the transitions from $^{11}$Li,
we can estimate the dissociation of $\alpha$ clusters in the daughter
 $^{11}$Be nucleus.
The observed strong GT transitions from $^{11}$Li to $^{11}$Be at 8 MeV
comes from the breaking of the one of the 2 $\alpha$ clusters in the excited
state of the daughter nucleus $^{11}$Be.

\end{document}